%% file: 00paper.tex
\documentclass[sigconf]{acmart}

\usepackage{booktabs}
\usepackage{algorithm}
\usepackage{algpseudocode}
\usepackage{bbold}

\usepackage{graphicx}
\usepackage{subcaption}

\newtheoremstyle{mydef}
{2ex}
{2ex}
{\itshape}
{}
{\scshape}
{: }
{0.5em}
{}
\theoremstyle{mydef}

\begin{document}

\copyrightyear{2018} 
\acmYear{2018} 
\setcopyright{acmcopyright}
\acmConference[SIGIR'18]{41st International ACM SIGIR Conference on Research and Development in Information Retrieval}{July 8--12, 2018}{Ann Arbor, MI, USA}
\acmBooktitle{SIGIR '18: 41st International ACM SIGIR Conference on Research and Development in Information Retrieval, July 8-12, 2018, Ann Arbor, MI, USA}
\acmPrice{15.00}
\acmDOI{10.1145/3209978.3210171}
\acmISBN{978-1-4503-5657-2/18/07}

\fancyhead{}

\title{SmartTable: A Spreadsheet Program with Intelligent Assistance}
\fancyhead{}

\author{Shuo Zhang}
\affiliation{%
  \institution{University of Stavanger}
}
\email{shuo.zhang@uis.no}

\author{Vugar Abdul Zada}
\affiliation{%
  \institution{University of Stavanger}
}
\email{v.abdulzada@stud.uis.no}

\author{Krisztian Balog}
\affiliation{%
  \institution{University of Stavanger}
}
\email{krisztian.balog@uis.no}

\begin{abstract}
We introduce SmartTable, an online spreadsheet application that is equipped with intelligent assistance capabilities.  With a focus on relational tables, describing entities along with their attributes, we offer assistance in two flavors: (i) for populating the table with additional entities (rows) and (ii) for extending it with additional entity attributes (columns).  We provide details of our implementation, which is also released as open source.  The application is available at \url{http://smarttable.cc}.
\end{abstract}



\maketitle

\input{00paper-01}

\input{00paper-02}

\input{00paper-03}
\input{00paper-04}
\input{00paper-05}
\input{00paper-06}

\bibliographystyle{ACM-Reference-Format}
\bibliography{00paper}

\end{document}

%% file: 00paper-01.tex
\section{Introduction}
\label{sec:int}

\begin{figure}[t]
   \centering
   \includegraphics[width=0.5\textwidth]{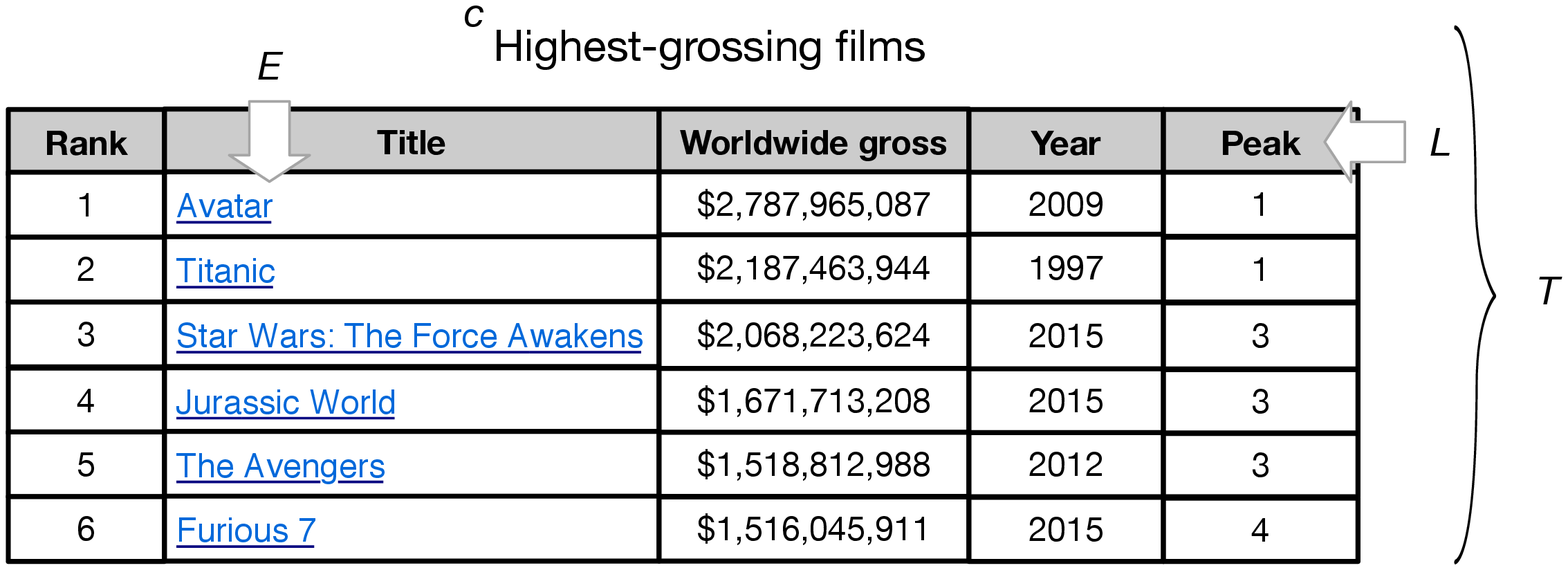} 
	\vspace*{-0.5\baselineskip}
   \caption{Example of a relational table $T$, where $c$ is the table caption, $E$ denotes the core column entities $E=\{e_1, \dots, e_n\}$, and $L$ is the set of column labels $L=\{l_1, \dots, l_m\}$.}    
	\vspace*{-\baselineskip}
	\label{fig:relational}
\end{figure}

Tables are a powerful, effective, and easy-to-use tool for both visual organization and manipulation of data.  Tables can be found in vast quantities on the Web, and spreadsheet programs are among the most commonly used desktop applications.
Our objective is to equip spreadsheet programs with intelligence assistance capabilities, to aid users while working with tables.
In this paper, we focus on one particular type of tables, known as \emph{relational tables}~\citep{Wang:2002:MLB,Wang:2002:DTH}.  Relational tables describe a set of entities along with their attributes.  We shall refer to the column containing the entities as the \emph{core column}.  Typically, it is either the leftmost table column or the second column from the left, in case row sequence numbering is used.  The heading labels of the table refer to particular attributes, while data cells hold the values of those attributes.  It is also assumed that the table is given a title (caption).  See Figure~\ref{fig:relational} for an illustration.


There exists a number of online tools and resources for table-related tasks, such as table search (Google Fusion Tables~\citep{Cafarella:2009:DIR} and WikiTables~\cite{Bhagavatula:2015:TEL}), question answering~\citep{Pasupat:2015:CSP}, and entity linking in tables~\citep{Bhagavatula:2015:TEL}. 
To the best of our knowledge, our system, called \emph{SmartTable}, is the first online spreadsheet program that provides intelligent table content recommendation.
Specifically, our application is capable of providing two kinds of assistance: (i) recommending additional entities, from an underlying knowledge base, to be added to the core column (row population) and (ii) recommending additional entity attributes to be included as columns (column population).
Such recommendations are particularly useful in scenarios with an exploratory or recall-oriented nature, i.e., when the user does not have a very clear idea beforehand as to what should be included in the table.
Additionally, SmartTable also provides regular table operations, such as adding, deleting, and moving rows and columns, editing cells, and supporting various value types (entities, numbers, currencies, dates, etc.).

Both types of assistance, that is, row and column population, are based on probabilistic models that we developed in prior work~\citep{Zhang:2017:ESA}. 
The main contributions of this paper are twofold. First, we integrate the above assistance functionality into an online spreadsheet application.
Second, we describe the task-specific indexing structures employed, and evaluate the efficiency of our implementation in terms of response time.
SmartTable is implemented using a HTML5 front-end and a Python+ElasticSearch back-end.  It uses DBpedia as the underlying knowledge base and a corpus of 1.6M tables extracted from Wikipedia.
The implementation is made open source at \url{https://github.com/iai-group/SmartTable} and the application is available online at \url{http://smarttable.cc}.


%

%% file: 00paper-02.tex
\section{Overview}
\label{sec:ove}

In this section, we provide an overview of the functionality of the SmartTable application, by walking through the process of creating a table from scratch. 

\begin{figure}[t]
        \centering
        \begin{subfigure}{\linewidth}
            \includegraphics[width=0.663\textwidth]{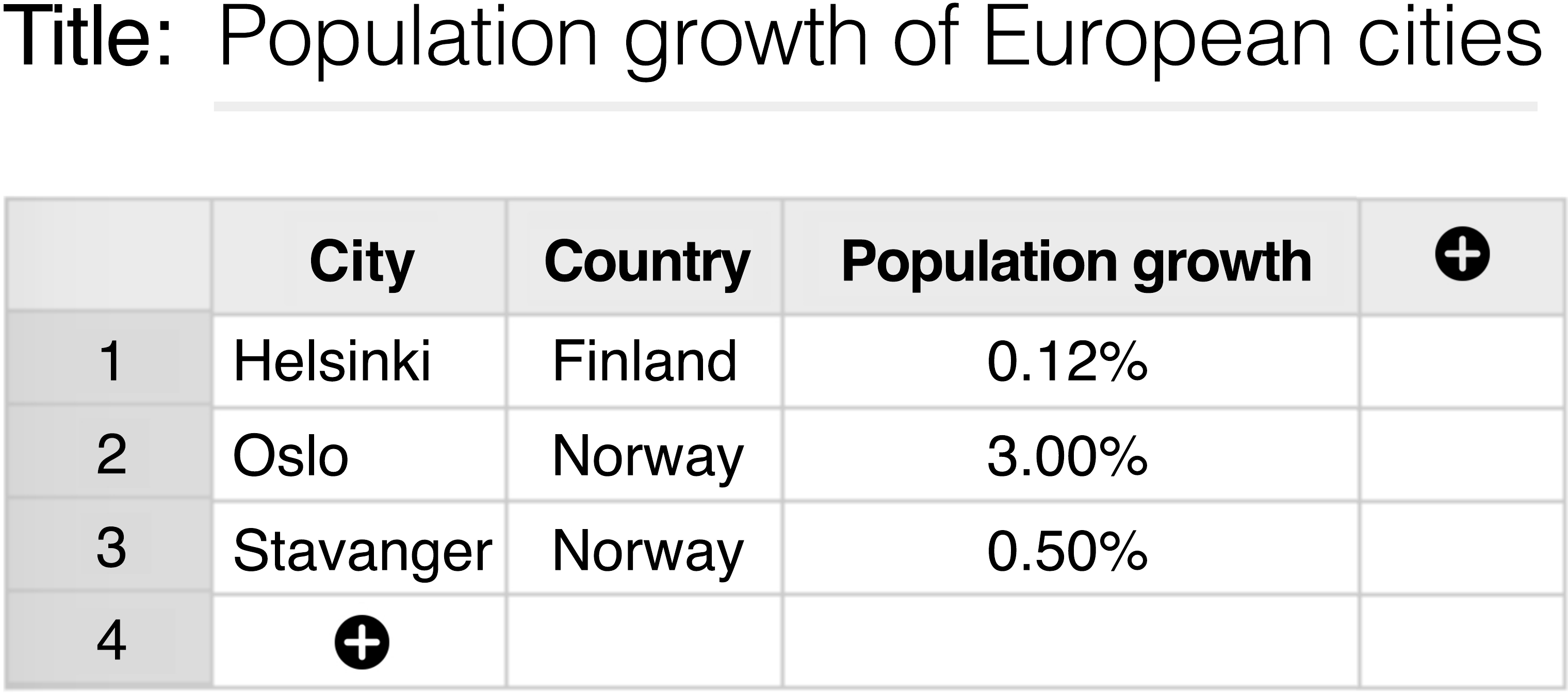}
            \caption{Seed table with some initial data.}
            \vspace*{\baselineskip}
            \label{fig:init}
        \end{subfigure}
        \hfill
        \begin{subfigure}{\linewidth}
            \includegraphics[width=0.663\textwidth]{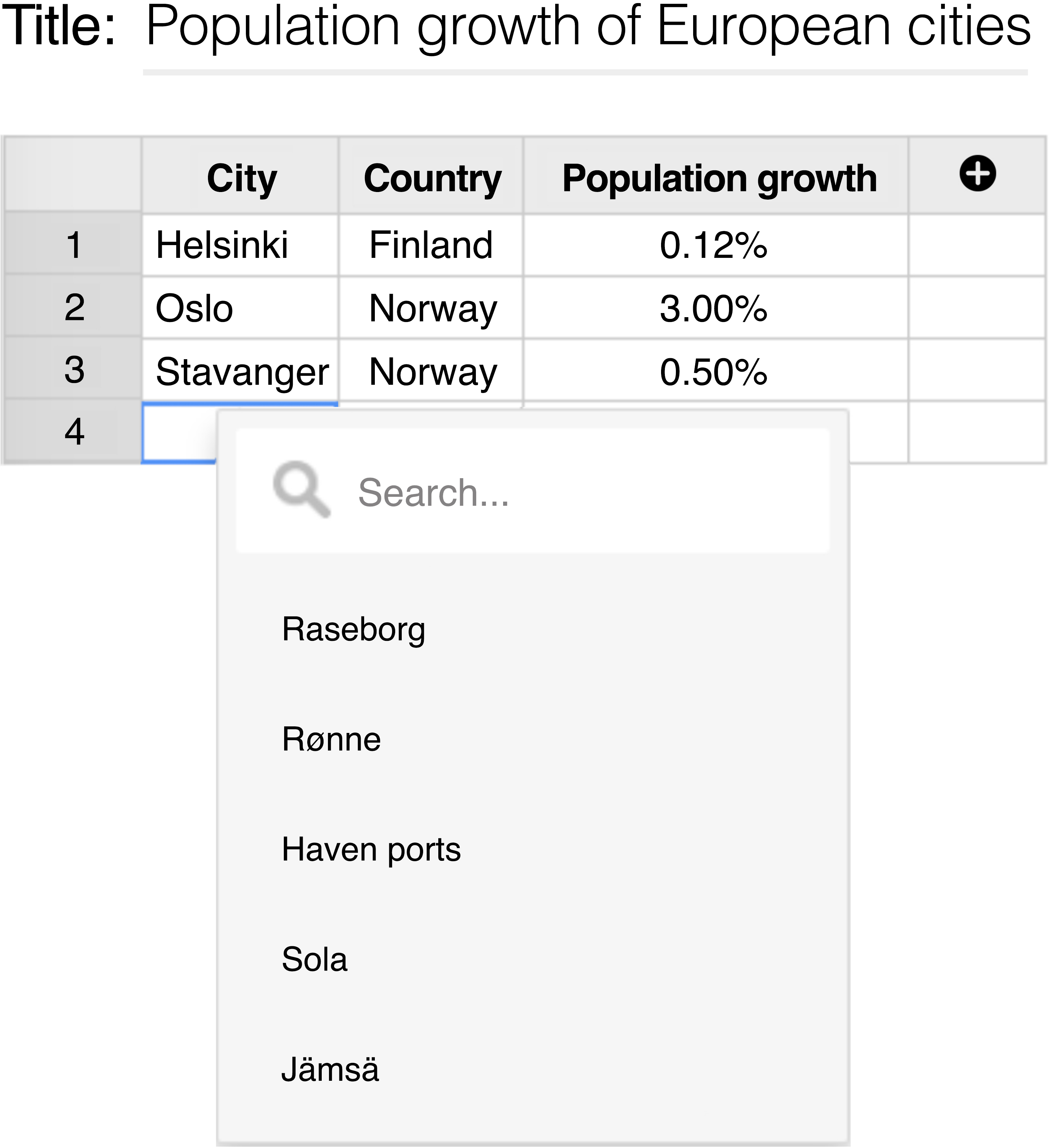}
            \caption{Row population assistance.}
            \label{fig:row}
            \vspace*{\baselineskip}
            \begin{minipage}{.1cm}
            \vfill
            \end{minipage}
        \end{subfigure} 
        \hfill
        \begin{subfigure}{\linewidth}
            \includegraphics[width=\textwidth]{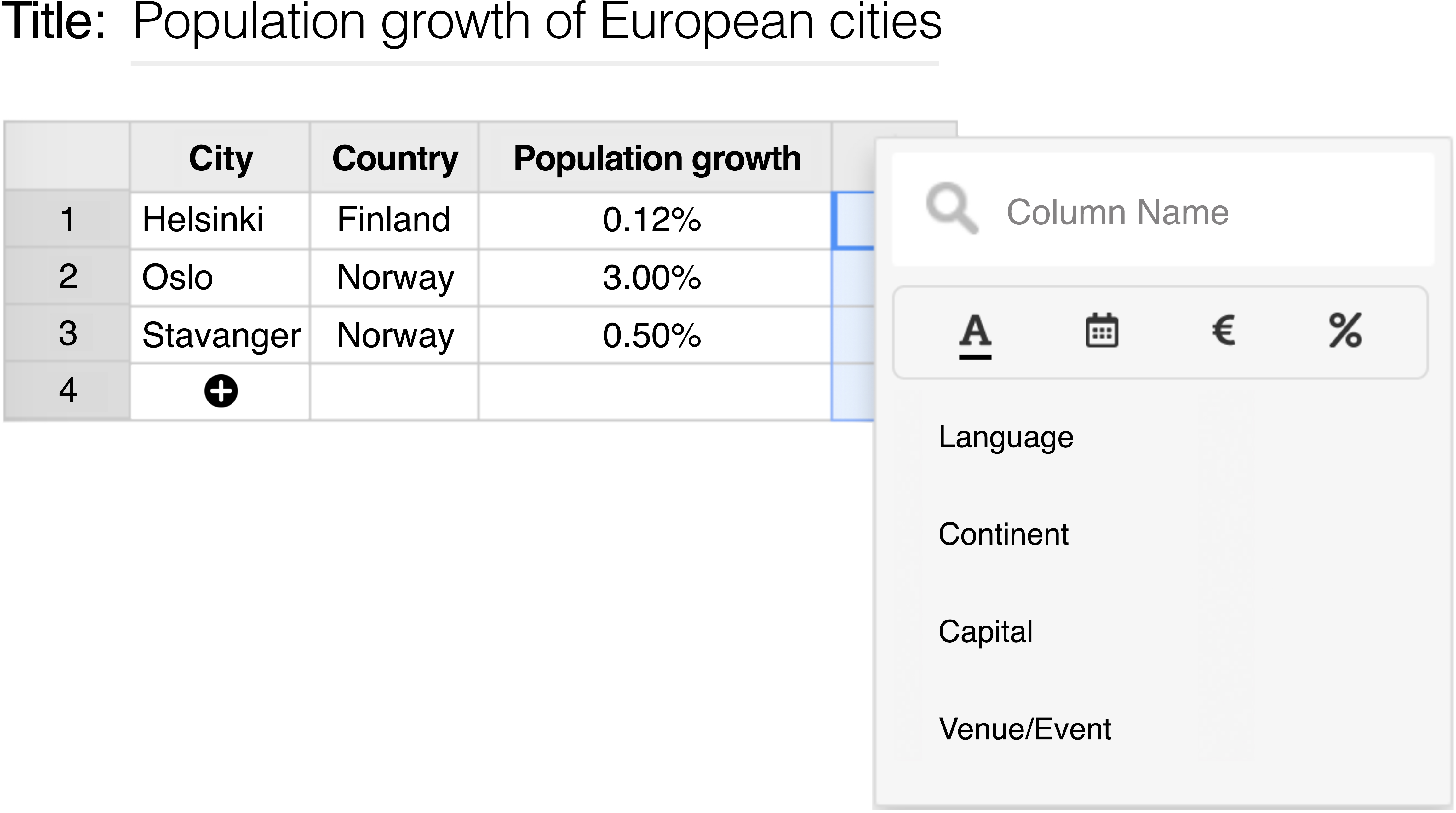}
            \caption{Column population assistance.}
            \label{fig:col}
            \begin{minipage}{.1cm}
            \vfill
            \end{minipage}
        \end{subfigure} 
        \caption{Screenshots from the SmartTable system.}
\end{figure}

\begin{itemize}
	\item Initially, we start with an empty table, with the table caption, core column entities, column labels, and cell values waiting to be filled. The user is expected to add a few entities and column labels first, along with an optional table caption, to supply the system with some data to base recommendations on.  We shall refer to this (incomplete) table as the \emph{seed table}. See Fig.~\ref{fig:init}.
	\item When adding entities to the core column, the user is presented with a ranked list of suggestions.  Additionally, the user can search the underlying knowledge base for entities.  See Fig.~\ref{fig:row}.
	\item When adding new columns, the user needs to specify the data type for that column (which can be one of entity, text, date, number, currency, or percentage) and provide a label for that column.  For the latter, a ranked list of suggestions are offered, along with a search box to search for additional labels.  See Fig.~\ref{fig:col}.
\end{itemize}

%% file: 00paper-03.tex
\section{Methods}
\label{sec:met}

In this section, we introduce the methods underlying the assistance functionality.  We refer to Fig.~\ref{fig:relational} for the notation used for the various table elements.  As for the data, we employ a table corpus (TC) extracted from Wikipedia and use DBpedia as the knowledge base (KB); further details about the datasets are given in Sect.~\ref{sub:sec:dat}.

\subsection{Row population}

Row population is the task of generating a ranked list of entities to be added to the core column of a given seed relational table.  It is closely related to the task of entity set expansion~\citep{Bron:2013:EBE,He:2011:SEI, Wang:2015:CEU}, which is about expanding a seed entity set with additional instances.
The main difference between row population and entity entity set expansion is that we can also leverage additional data from the seed table as input, not only the core column entities.  The row population task is split into two sub-tasks, which are candidate selection and ranking entities, respectively.

\subsubsection{Candidate selection}

We identify candidate entities using both the knowledge base and the table corpus.  From the knowledge base, we take entities that share the assigned semantic categories with those of the seed entities.  From the table corpus, we first find tables similar to the seed table, based on table caption, core column entities, and column heading labels. Then, we take the core column entities from those similar tables as candidates.

\subsubsection{Ranking entities}
\label{sec:sub:re}
We implement the probabilistic model proposed in~\citep{Zhang:2017:ESA}, which is a multi-conditional probability:
\begin{align*}
	P(e|E,L,c) & \propto P(e|E)P(L|e)P(c|e) ~, \label{eq:joint}
\end{align*}	
where $P(e|E)$ is entity similarity, $P(L|e)$ denotes column labels likelihood, and $P(c|e)$ is caption likelihood. 

\begin{description}
	\item[Entity similarity] is estimated using 
	\begin{equation*}
		P(e|E) = \lambda_E P_{KB}(e|E) + (1-\lambda_E) P_{TC}(e|E) ~,
	\end{equation*}
	where $P_{KB}(e|E)$ is the average Jaccard similarity between the candidate entity $e$ and each seed entity $e' \in E$, and $P_{TC}(e|E)$ is fraction of tables in the table corpus that contain both the seed and candidate entities out of the number of tables containing the seed entities.
	\item[Column labels likelihood] considers the table corpus and is estimated using a Dirichlet-smoothed language model:\footnote{In our original approach~\citep{Zhang:2017:ESA} this estimate was a two-component mixture. Due to efficiency considerations, we use a simplified version here. The relative difference in terms of effectiveness is below 5\%.} 
	\begin{equation*}
		P(L|e)=\sum_{l \in L}\prod_{t \in l} \frac{tf(t,e)+\mu P(t|\theta)}{|e|+\mu} ~,
		\label{eq:lm}
	\end{equation*}
	where $tf(t,e)$ is the term frequency of $t$ in the column labels of tables containing $e$ and $|e|$ is the sum of all term frequencies for $e$. The collection language model $P(t|\theta)$ is computed based on the column labels of all tables in TC.
	\item[Caption likelihood] is a two-component mixture: 
	\begin{equation*}
		P(c|e) = \prod_{t \in c} \big( \lambda_c P_{KB}(t|\theta_e) + (1-\lambda_c) P_{TC}(t|e) \big ) ~,
	\end{equation*}
	where the KB component is estimated using a Dirichlet-smoothed entity language model.  The TC component is computed as $P_{TC}(t|e)=\#(t,e)/\#(e)$, where $\#(t, e)$ is the number of tables in the table corpus containing entity $e$ in the core column and term $t$ in the table caption, and $\#(e)$ is the total number of tables contaning $e$.
\end{description}
%
%
%
%
%

%

\subsection{Column population}
Column population is the task of generating a ranked list of column labels to be added to the column headings of a given seed table.  It is also implemented as a sequence of two steps: candidate selection and column label ranking. 

\subsubsection{Candidate selection}
Candidate labels are obtained from related tables. 
To find related tables, we use (i) the table caption, (ii) table entities, and (iii) seed column heading labels as queries.  From the matching tables, column labels are extracted as candidates. 

\subsubsection{Ranking column labels}
The related tables, identified in the candidate selection stage, are also utilized in the ranking step.  According to the model in~\citep{Zhang:2017:ESA}, the probability of a candidate column label is given by:
\begin{equation*}
	P(l|E,c,L) = \sum_T P(l|T) P(T|E,c,L) ~,
\label{eq:ptecl}
\end{equation*}
where $T$ represents a related table, $P(l|T)$ is the label's likelihood given $T$, and $P(T|E,c,L)$ expresses that table's relevance. 
The probability $P(l|T)$ is set to 1 if the candidate label $l$ is present in table $T$ and is 0 otherwise. 
The relevance of a table is estimated as:
\begin{equation*}
	P(T|E,c,L) \propto P(T|E)P(T|c)P(T|L) ~,
\end{equation*}
where $P(T|E)$ denotes entity coverage, $P(T|c)$ is caption likelihood, and $P(T|L)$ is the column labels likelihood.
\begin{description}
	\item[Entity coverage] is the overlap of core column entities in the seed table and in $T$: $P(T|E) = |T_E \cap E|/|E|$. 
	\item[Caption likelihood] is estimated using term-based similarity between the seed table's caption and the content of $T$: 
	$P(T|c) \propto sim(T_c, c)$. Here, we employ BM25 scoring.
	\item[Column labels likelihood] is the overlap between labels of the seed table and those of $T$: $P(T|L) = |T_L \cap L|/|L|$.
\end{description}
%


%% file: 00paper-04.tex
\section{Implementation}
\label{sec:imp}

In this section, we describe the datasets used and indices built, along with technical details of our implementation.

\subsection{Datasets}
\label{sub:sec:dat}

We rely on two data sources: a table corpus and a knowledge base.
The knowledge base is DBpedia, version 2015-10.\footnote{\url{http://wiki.dbpedia.org/dbpedia-dataset-version-2015-10}}  We filter out entities that do not have a short textual description (abstract).  After filtering, we are left with a total of 4.6M entities.
As for the table corpus, we use the WikiTables collection~\cite{Bhagavatula:2015:TEL}, which comprises of 1.65M tables, extracted from Wikipedia. We preprocess tables as follows.  Entities are marked up in the original table with hyperlinks.  If the link points to an entity that exists in DBpedia, we replace that link with the corresponding entity identifier.  Otherwise, we replace the link with the anchor text.

\subsection{Indices} 
\label{sub:sec:ds}

\begin{figure}[t]
   \centering
   \includegraphics[width=0.25\textwidth]{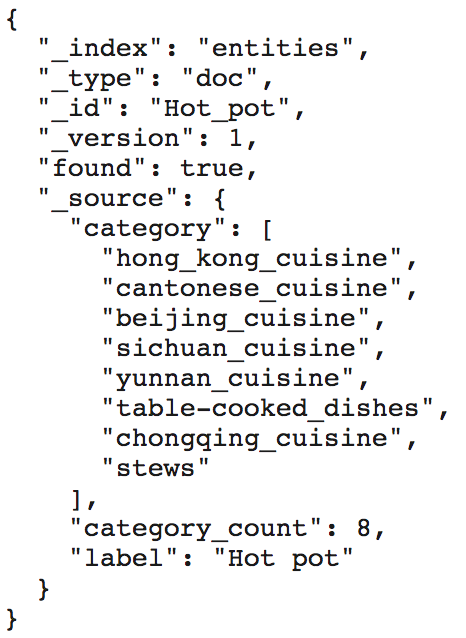} 
   \caption{Example entry from the entity index.}    
	\label{fig:index_example}
\end{figure}

We build the following inverted indices:


%
\begin{description}
	\item[Table index] It contains 1.65M Wikipedia tables (6.4GB).  For each table, the following fields are stored: page title, section title, table caption, column labels, table data, and core column entities.
	\item[Entities] It contains 4.6M DBpedia entities (2GB).  For each entity, we store its canonical name (label), and the list and number of categories it is assigned to.  See Fig.~\ref{fig:index_example} for an example.
	\item[Categories] We use Wikipedia's category system, comprising of around 1M categories. For each category, we store the list of entities that are assigned to that category. This index occupies 2GB. 
\end{description}

\subsection{Implementation}
SmartTable is a web application that is comprised of a HTML5 front-end and a back-end based on Python and Elasticsearch.

\subsubsection{Front-end}

The front-end stack is made up of HTML, CSS, and JavaScript (ECMAScript6 standard). 
We build on a third-party JavaScript spreadsheet framework called Handsontable,\footnote{\url{https://handsontable.com/}} which provides a rich set of functionality for tables, including sorting, conditional formatting, contextual menus, moveable and resizable rows and column, etc.
Additionally, we utilize the \texttt{Gulp.js}, \texttt{Babel.js}, and \texttt{Node.js} JavaScript libraries.

For development, we follow the MVC (Model View Controller) software architecture pattern. 
The system is divided into self-conta\-ined components that are easy to debug and maintain, with loose coupling and modularity between the fundamental parts.  Figure~\ref{fig:fro} provides an overview.  At the center of front-end lies the TableContainer class, connecting the following components:

\begin{figure}[t]
   \centering
   \includegraphics[width=0.5\textwidth]{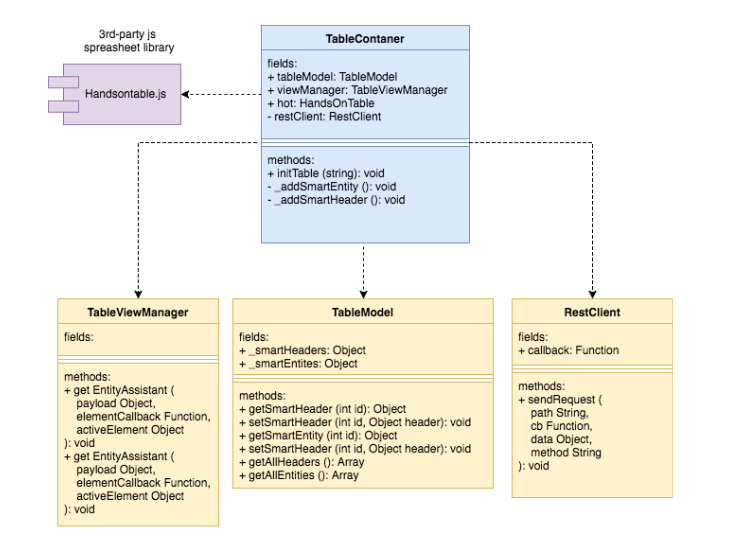} 
	\vspace*{-1\baselineskip}
   \caption{Overview of the application front-end.}    \label{fig:front-end}
	\vspace*{-\baselineskip}
	\label{fig:fro}
\end{figure}
\begin{itemize}
	\item Handontable.js: Third party JavaScript spreadsheet framework. 
	\item TableViewManager.js: Smart Assistant view controller.
	\item TableModel.js: Provides storage and accessibility to all core column entities and column heading labels.
	\item RestClient.js: Communication component, which is responsible for request sending and response provision via the respective callback calls.
\end{itemize}

\subsubsection{Back-end}

The back-end consists of two parts: a web ser\-ver and a recommendation engine. The main role of the former  is to connect the front-end spreadsheet application (client) with the recommendation engine.  
The web server is implemented in Python, using the Flask framework.\footnote{\url{http://flask.pocoo.org/}}
Communication is done over HTTP, with request and response messages encoded in JSON format.  
The recommendation engine is responsible for generating the ranked list of suggestions (entities and column labels).  It uses Elasticsearch as the underlying indexing and retrieval engine.  All indices are built using the Nordlys toolkit~\citep{Hasibi:2017:NTE}.\footnote{\url{http://nordlys.cc}} 

%% file: 00paper-05.tex
\section{Evaluation}
\label{sec:eva}

\begin{figure}[t] 
\centering
  \begin{subfigure}[t]{0.5\linewidth}
    \includegraphics[width=1\linewidth]{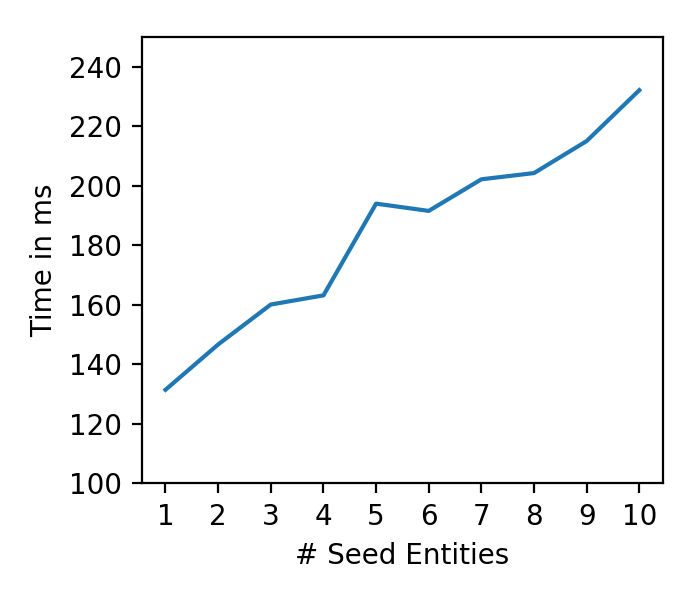} 
    \caption{Row population} 
    \label{fig:eval_row}
  \end{subfigure}
  \begin{subfigure}[t]{0.5\linewidth}
    \includegraphics[width=1\linewidth]{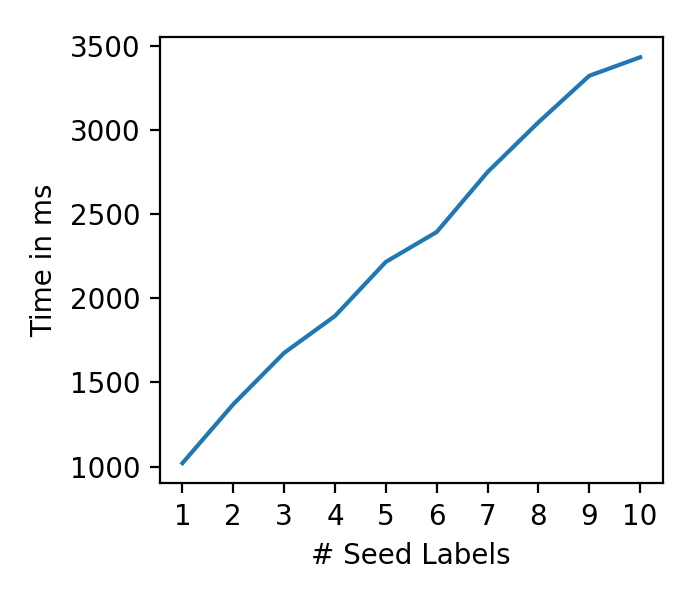} 
    \caption{Column population} 
    \label{fig:eval_col}
  \end{subfigure} 
  \caption{Performance in terms of response time.}
\end{figure}

In previous work~\citep{Zhang:2017:ESA}, we have performed an extensive evaluation of the row and column population methods in terms of effectiveness.
Here, we evaluate our system in terms of efficiency.  We measure response time as the time elapsed between receiving the request and sending off the response on the back-end, i.e., net computation time without the network overhead.
Using 10 random tables, we vary the number of core column entities (seed entities) and the number of heading column labels (seed labels).  The measurements are repeated 10 times and averages are reported in Figs.~\ref{fig:eval_row} and~\ref{fig:eval_col}.
We can observe that, in both cases, response time grows linearly with the size of the input.
For row population, the response time is beyond 250ms, even with the largest input size, which is considered very acceptable.
For column population, responses are a magnitude slower.  This is due to the fact that we consider all related tables in our scoring formula.  Limiting the computations to the top-$k$ most similar tables may provide a solution; it is left for future work to find a $k$ value that provides a good trade-off between effectiveness and efficiency.


%

%

%% file: 00paper-06.tex
\section{Conclusion and future work}
We have introduced SmartTable, an online spreadsheet application that is equipped with smart assistance capabilities.  Specifically, we aid users working with relational tables by suggesting them additional entities and column heading labels to be included in the table.  In future work, we consider diversifying recommendations and plan to extend the scope of content recommendation to data cells as well, by suggesting possible values for them.  Furthermore, we intend to integrate table search and table generation functionality, which we developed in recent work~\citep{Zhang:2018:AHT,Zhang:2018:OTG}.